\documentclass[aps,prl,twocolumn,superscriptaddress,showpacs]{revtex4}

\usepackage[dvips]{graphicx}
\usepackage{latexsym}
\usepackage{bm}

\begin{document}

\title{
Sign reversal of field-angle resolved heat capacity oscillations in a heavy fermion superconductor 
CeCoIn$_5$ and $d_{x^2-y^2}$ pairing symmetry
}

\author{
K. An
}
\affiliation{
Institute for Solid State Physics, University of Tokyo, Kashiwa
277-8581,
 Japan
}
\author{
T. Sakakibara
}
\affiliation{
Institute for Solid State Physics, University of Tokyo, Kashiwa 277-8581,
Japan
}
\author{
R. Settai
}
\affiliation{
Department of Physics, Osaka University, Toyonaka 560-0043, Japan
}
\author{
Y. Onuki
}
\affiliation{
Department of Physics, Osaka University, Toyonaka 560-0043, Japan
}
\author{
M. Hiragi
}
\affiliation{
Department of Physics, Okayama University, Okayama 700-8530, Japan
}\author{
M. Ichioka
}
\affiliation{
Department of Physics, Okayama University, Okayama 700-8530, Japan
}
\author{
K. Machida
}
\affiliation{
Department of Physics, Okayama University, Okayama 700-8530, Japan
}

\date{\today}

\begin{abstract}
To identify the superconducting gap symmetry in CeCoIn$_5$ ($T_{\rm c}$=2.3~K), 
we performed 
angle-resolved specific heat ($C_\phi$) measurements in a field rotated around the $c$-axis down to very low temperatures 0.05$T_{\rm c}$ and detailed theoretical calculations.
In a field of 1~T, a sign reversal of the fourfold angular oscillation in $C_\phi$ has been observed at $T\simeq 0.1T_{\rm c}$ on entering a quasiclassical regime
where the maximum of $C_\phi$ corresponds to the antinodal direction,
coinciding with the angle-resolved density of states (ADOS) calculation.
The $C_\phi$ behavior, which exhibits minima along [110] directions,
unambiguously allows us to conclude $d_{x^2-y^2}$ symmetry of this system.
The ADOS-quasiclassical region is confined to a narrow $T$ and $H$ domain within 
$T/T_{\rm c}\sim 0.1$ and 1.5~T (0.13$H_{\rm c2}$).

\end{abstract}


\maketitle

Investigation of superconductivity in strongly correlated electron systems has been a fascinating subject over the past few decades, since a realization of unconventional non
$s$-wave pairings can be expected due to strong electron-electron repulsion.
A gap function of these unconventional superconductors (SCs) mostly has zeros (has nodes) along certain directions in the momentum space.
The existence of nodes (point or line) can be inferred from power-law dependences in the physical quantities such as the specific heat ($C$) or the nuclear spin relaxation rate.
To identify the gap structure is, however, a more formidable task. For instance, $d_{xy}$ and $d_{x^2-y^2}$ symmetry differ from each other only in the position of line nodes (45$^{\circ}$ rotation of the latter becomes identical with the former). 
Phase-sensitive or direction-sensitive experiments are therefore needed to discriminate these gap structures.
Very elegant phase-sensitive experiments have been done on the high-$T_{\rm c}$ cuprates to establish pure $d_{x^2-y^2}$ symmetry,  employing sophisticated tricrystal or tetracrystal geometries~\cite{tsuei}.

Here we address the directional thermodynamic measurements.
These techniques utilize that the low-lying quasiparticle excitations near the gap nodes become field  orientation dependent~\cite{vekhter,miranovic} in a low $H$ and  $T$ region
where the nodal quasiparticles contribute predominantly
to thermodynamic quantities~\cite{volovik};
The angle-resolved density of states (ADOS) $N_0({\phi})$ at the Fermi energy 
becomes the largest (smallest) when $H$ is parallel to the antinodal (nodal) directions. 
The nodal directions can be determined from the angular variation of $C_{\phi}$ or the thermal conductivity $\kappa_{\phi}$ in a rotating $H$ (minima of $C_{\phi}$ or $\kappa_{\phi}$ occur along the nodal directions).
Thus those angle-resolved bulk thermodynamic experiments provide
an indispensable spectroscopic method to locate the nodal direction.

Up to now, the $C_{\phi}$ experiments have been performed on various anisotropic SCs to probe the gap structures~\cite{t-park03,aoki,deguchi,yamada,yano,t-park08}.
Among these, we concentrate here on the heavy fermion superconductor CeCoIn$_5$ ($T_{\rm c}$=2.3~K) in which the gap structure has been controversial.
The previous $C_{\phi}$ experiment ($H$ rotated in the $ab$-plane) performed down to 0.3~K ($T/T_{\rm c}\geq$0.13) revealed a fourfold angular oscillation with the minima along [100], from which $d_{xy}$ symmetry was deduced~\cite{aoki}. 
The same symmetry was also inferred from the anisotropy in the high-field superconducting phase of CeCoIn$_5$~\cite{r-ikeda}.
This conclusion was, however,  in disagreement with that of the $\kappa_{\phi}$ measurements in which $d_{x^2-y^2}$ symmetry was derived~\cite{izawa}. 
Moreover, instability of the vortex lattice structure observed by small-angle neutron scattering experiments~\cite{bianchi,kawamura} has been shown to be consistent with the $d_{x^2-y^2}$ pairing~\cite{hiasa,machida}.  
Point-contact Andreev reflection experiment~\cite{w-park} and the spin resonance observed in the superconducting state~\cite{stock,eremin} also support 
$d_{x^2-y^2}$ symmetry in CeCoIn$_5$.

Recently a resolution of this discrepancy has been put forward by Vekhter and co-workers~\cite{vorontsov,boyd}. 
They theoretically examined the phase diagram for $C_{\phi}$ of $d$-wave SCs in a wide $H$-$T$ region 
and pointed out that applying $H$ along the gap nodes may result in maxima of $C_{\phi}$ in an intermediate-$T$ range.
It is due to an inversion of the anisotropy in the finite energy DOS when the scattering of the quasiparticles off the vortices is considered. 
Thus, the ADOS-quasiclassical region in which $C_{\phi}$ takes minima for nodal direction may lie at still lower temperatures. 
Such a crossover in $C_{\phi}$ with $T$ has not yet been observed in nodal SCs.
If this is the case for CeCoIn$_5$, then the anisotropy of $C_{\phi}$ is expected to change the sign at a temperature $\simeq 0.1T_{\rm c}$~\cite{vorontsov,boyd}. 
In order to explore this possibility, we have extended the $C_{\phi}$ measurements on CeCoIn$_5$ down to 0.05$T_{\rm c}$. Here we also perform more accurate calculations 
$C_{\phi}$ without resorting to the Pesch approximation 
adopted by Vorontsov {\it et al.}~\cite{vorontsov} and we also take into account Pauli paramagnetic effect and Fermi velocity anisotropy
effect, both of which are known to be important in this system~\cite{bianchi,hiasa}.
Those experimental and theoretical combined efforts pave a way to
establish the angle-resolved thermodynamic measurements $C_{\phi}$ and 
$\kappa_{\phi}$ as important spectroscopic means to determine the
pairing symmetry of unconventional superconductors.

The single crystal of CeCoIn$_5$ (23~mg weight) used in the present experiment was a thin plate whose $c$ axis oriented perpendicular to the largest face. 
$C_{\phi}$ measurements were done by a semiadiabatic heat pulse method. 
The sample was cooled using a dilution refrigerator (Oxford Kelvinox AST Minisorb) with a double sorption pump system built inside the insert (51~mm outer diameter) to allow continuous circulation of $^3$He. 
Having no thick pumping tube outside, the insert could be easily rotated using a stepper motor mounted at the top of a magnet Dewar.
Horizontal fields were generated by a split-pair superconducting magnet.
We confirmed that the addenda contribution, which was always less than 10\% of the sample specific heat, has no field-angle dependence.
In a fixed field strength, we took the $C_\phi$ data every 2 degrees of the $H$ rotation in the $ab$-plane covering [100] through [010] directions.
For each angle, the $C_\phi$ value was determined by the average of three successive measurements. 

\begin{figure}[t]
\includegraphics[width=8.cm]{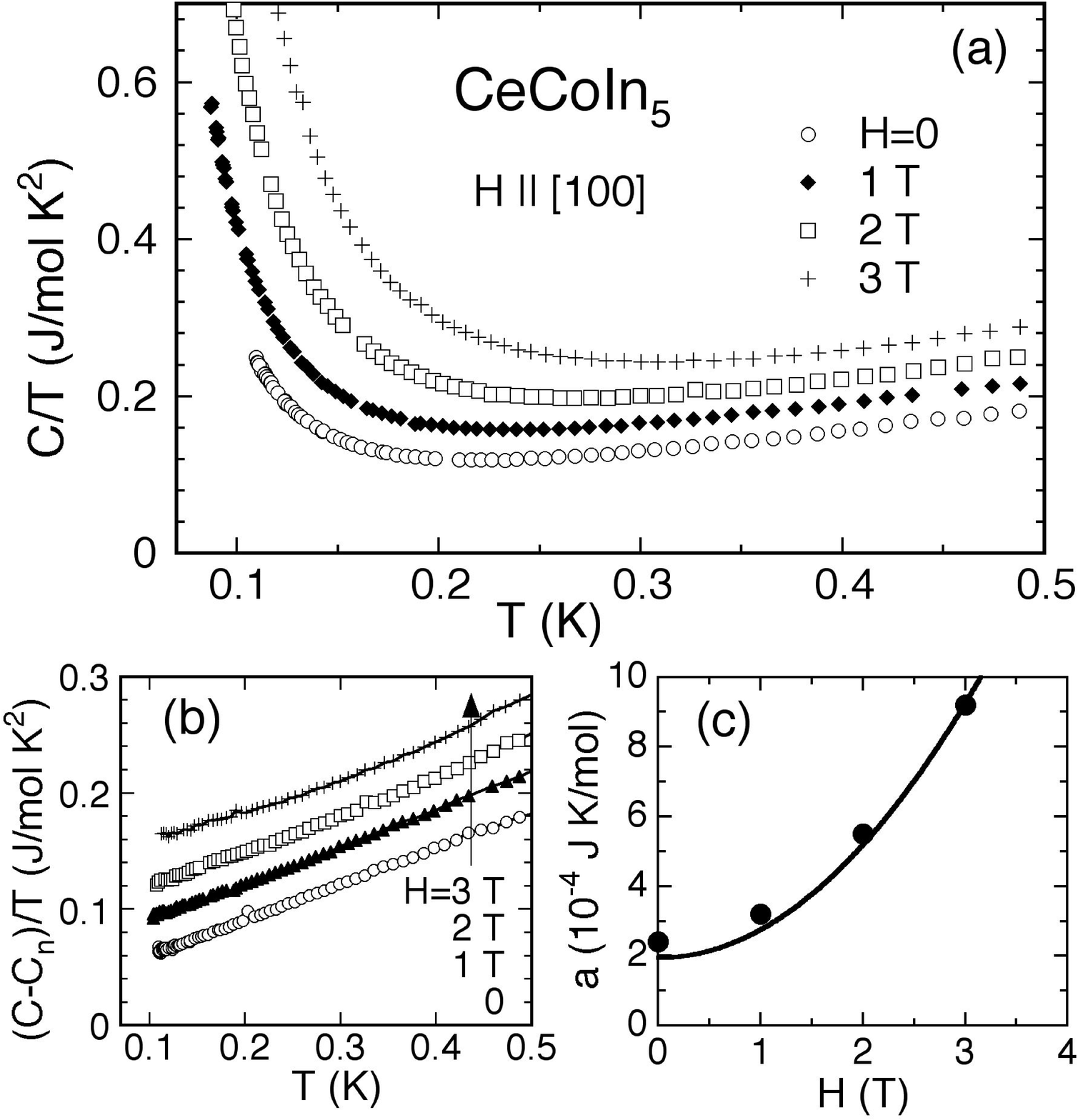}
  \caption{
(a) $C/T$ of CeCoIn$_5$ at low temperatures measured in various fields applied along the [100] direction.
(b) The electronic part $(C-C_{\rm n})/T$ obtained by subtracting the nuclear contribution.
(c) Field variation of the coefficient $a(H)$ of the nuclear Schottky term.
Dots and the solid line are the the experimental and the calculated results, respectively.
}
 \label{fig:fig1}
\end{figure}

First we show in Fig.~1(a) the temperature variation of $C/T$ obtained at $H$=0, 1, 2 and 3~T in a temperature range below 500~mK.
At zero field, $C/T$ above 300~mK is linear to $T$ as expected for the line node SC.  
At low temperatures below 200~mK, an upturn in $C/T$ shows up. 
This upturn is due to a quadrupole splitting of the $^{115}$In ($I$=9/2) and $^{59}$Co ($I$=7/2) nuclear spins~\cite{movshovich,s-ikeda}. 
The nuclear contribution significantly increases by applying $H$ and dominates the $C/T$ data below 0.3~K.
We subtracted the nuclear Schottky contribution $C_{\rm n}=a(H)/T^2$, where the coefficient $a(H)$ is determined at each field so that the resulting electronic contribution $(C-C_{\rm n})/T$ becomes linear to $T$ at low temperatures as shown in Fig.~1(b)~\cite{nakai}. 
Dots in Fig.~1(c) denote $a(H)$ obtained by this analysis, which show a $H^2$ like increase due to a nuclear Zeeman splitting.
Of course, the above evaluation of $a(H)$ might have some arbitrariness.
Fortunately, the manner of subtracting $C_{\rm n}$ does not strongly affect the analysis of $C_\phi$ below, since $C_{\rm n}$ is considered to be $\phi$-independent as will be discussed later.
The rapid increase of $C_{\rm n}$ at low $T$, however, sets the lower bound of the temperature for the present $C_\phi$ measurements to be $\sim$0.12~K for $H$=1~T and $T>$0.16~K for $H$=2 and 3~T, because of a small amplitude of the oscillation and a finite resolution ($\sim$0.5~\%) of the specific heat measurements.

\begin{figure}[t]
\includegraphics[width=8.5cm]{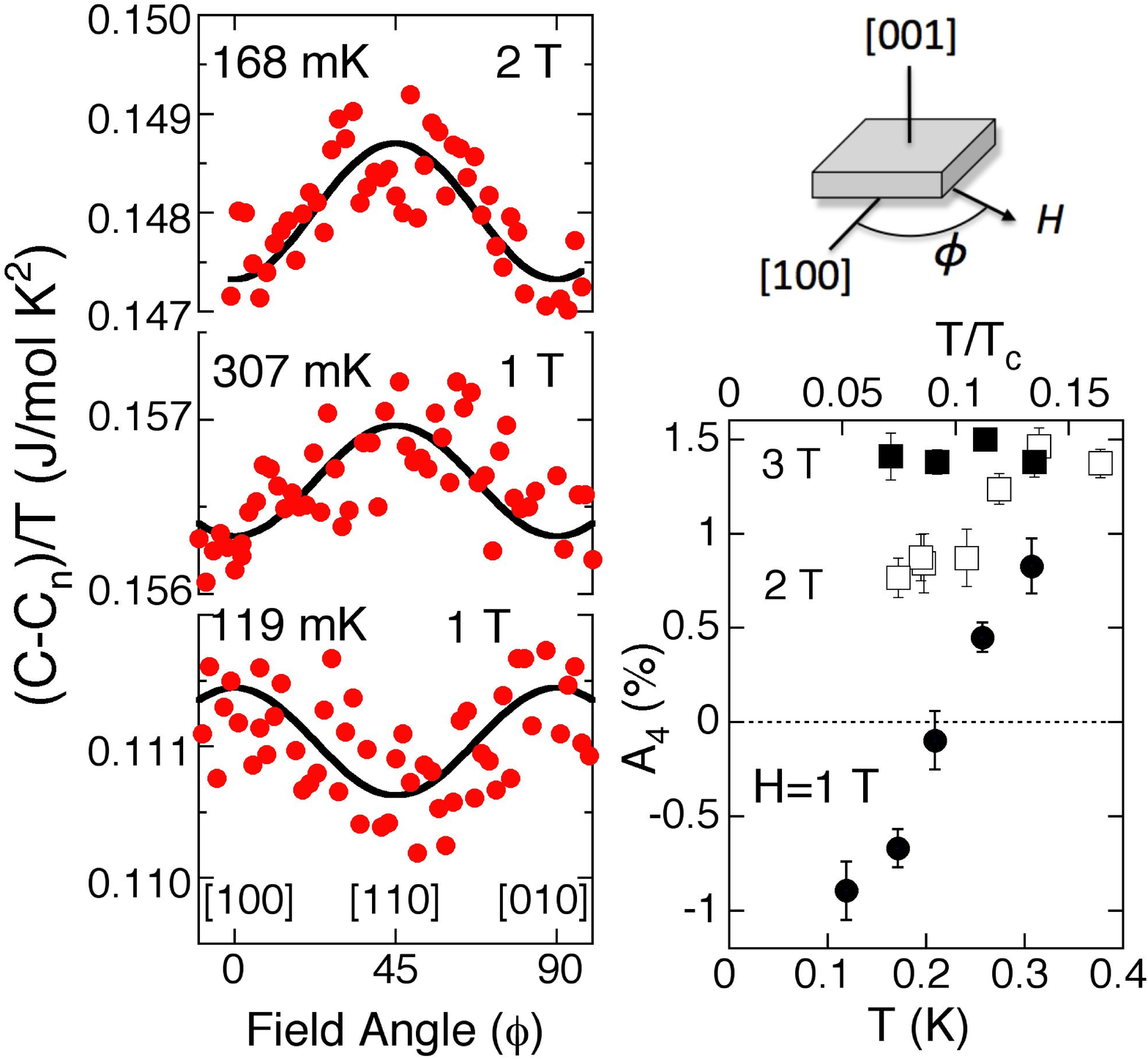}
  \caption{(color online).
Left panel: Angular dependent specific heat data for selected fields and temperatures. 
Magnetic field is rotated in the $ab$-plane. 
Solid lines are fits to the expression $C_0$+$C_H(1-A_4\cos 4\phi$).
Right panel: Temperature variation of the normalized fourfold amplitude $A_4$ obtained for $H=$3, 2 and 1~T.
}
 \label{fig:fig2}
\end{figure}

In Fig.~2 (left panel), we show the nuclear subtracted $C_\phi$ data for selected temperatures and fields. 
We always observed a fourfold oscillation of $C_\phi$ with no appreciable twofold part. 
We fit the data to an expression
$C-C_{\rm n}=C_0+C_H(1-A_4\cos 4\phi)$, where $C_0$ and $C_H$ are the zero field and field dependent parts of the electronic specific heat, respectively, and $\phi$ is the angle of $H$ measured from [100] direction. 
$A_4$ is the normalized amplitude of the fourfold oscillation which, for nodal SCs, becomes proportional to the zero-energy DOS anisotropy $|A_4|\propto 1-N_0(H\!\parallel\! {\rm node})/N_0(H\!\parallel\! {\rm antinode})$ in an ADOS-quasiclassical regime at low $T$ and takes a nonzero value as $H\rightarrow 0$~\cite{miranovic}.
The $A_4$ value thus obtained is shown in Fig.~2 (right panel) as a function of $T$ for $H$=1, 2 and 3~T.
At $T$=300~mK, $A_4$ is positive ($C_\phi$ minima along [100]) for all fields in agreement with the previous measurements~\cite{aoki}. 
For $H$=3~T, $A_4$ is nearly $T$-independent below 300~mK.  
$A_4$ for $H$=2~T shows a slight decrease on cooling, although it appears to remain positive as $T\rightarrow$0. 
When $H$ is reduced to 1~T, $A_4$ changes the sign around 200~mK and becomes {\it negative} ($C_\phi$ minima along [110]) at lower temperatures.

In the evaluation of $A_4$ above, we assumed that $C_{\rm n}(H)$ is isotropic in the $ab$-plane. 
We confirmed the validity of this assumption by calculating $C_{\rm n}(H)$ from the nuclear spin Hamiltonian~\cite{curro}
$\mathcal{H}_{\rm n}=(h\nu_Q/6)\bigl[3I_z^2-I(I+1)+\eta(I_x^2-I_y^2)\bigr]+\gamma\hbar\bm{I}(\bm{1}+\tilde{\bm{K}})\bm{H}$,
where $\nu_Q$ and $\eta$ are the parameters describing a quadrupole splitting, and $\tilde{\bm{K}}$ is the Knight shift tensor.
In CeCoIn$_5$, there are four low-symmetry ($\eta\neq 0$) In sites (per unit cell) on the lateral faces of the unit cell.
$\tilde{\bm{K}}$ also has an in-plane anisotropy in these In sites.
Each site of these then produces a twofold in-plane anisotropy in $C_{\rm n}(H)$. 
When averaged over the four sites, however, the twofold anisotropy is cancelled, leaving no (fourfold) $\phi$ dependence in $C_{\rm n}(H)$.
The solid line in Fig.~1(c) is the field variation of $a(H)$ calculated with the parameters taken from Ref.~\cite{curro}, which is in good agreement with the experimentally evaluated results.
We may therefore conclude that the behavior of $A_4(T,H)$ above is the intrinsic properties in the superconducting state.

\begin{figure}[t]
\includegraphics[width=7cm, height=6cm]{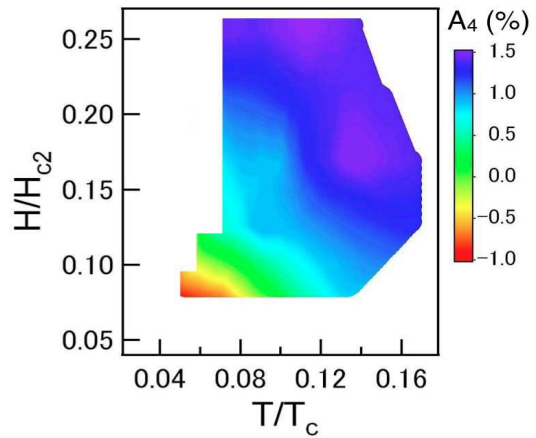}
  \caption{
 Contour plot of the normalized fourfold amplitude $A_4(T,H)$ of CeCoIn$_5$ obtained by the present experiment. $T$ and $H$ are normalized by $T_{\rm c}$=2.3~K and $H_{\rm c2}$=11.5~T, respectively.
 }
 \label{fig:fig3}
\end{figure}

In Fig.~3, we show the contour plot of $A_4(T,H)$ obtained by the present measurements.
Our present results are fully compatible with the previous $C_{\phi}$ experiment performed in the region $T\geq 0.13T_{\rm c}$ and $H\geq 0.17H_{\rm c2}$, where  $A_4(T,H)$ is positive and $C_{\phi}$ takes minima along [100] directions.
The sign reversal in $A_4(T,H)$ is observed only at much lower temperatures ($T\leq 0.12T_{\rm c}$) and lower fields ($H\leq 0.15H_{\rm c2}$).
As will be shown in the following, the crossover temperature of the sign reversal ($\sim$0.12$T_{\rm c}$) well agrees with the theoretical prediction. 
These observations indicate that,  for the first time, the low temperature ADOS-quasiclassical region is reached experimentally. 
In this region, $C_{\phi}$ takes minima along [110] directions implying unambiguously that the gap symmetry of CeCoIn$_5$ is $d_{x^2-y^2}$.

In order to understand the above data of $C_{\phi}$ or $A_4$
and also to help establishing the angle-resolved thermodynamic
measurements, we calculate the angle-resolved specific heat 
$C_{\phi} (T,H)$ accurately in the quasiclassical Eilenberger theory
without resorting to the Pesch approximation 
adopted by Vorontsov {\it et al.}~\cite{vorontsov}.
The Eilenberger equations are read as 
\begin{eqnarray}
\{\omega_n+i\mu B+\tilde{{\bf v}}_F\cdot[\nabla+i{\bf A}(r)]\}f=\Delta(r)\bar{\phi}(k)g \\
\{\omega_n+i\mu B-\tilde{{\bf v}}_F\cdot[\nabla-i{\bf A}(r)]\}f^{\dagger}=
\Delta^{\ast}(r)\bar{\phi}(k)g
\end{eqnarray}

\noindent
where the quasiclassical Green functions are related to $g=(1-ff^{\dagger})^{1/2}$.
The $d_{x^2-y^2}$ pairing function  is given by $\bar{\phi}(k)=\cos2\theta$
where $\theta$ is azimuthal angle in $k$-space.
The Pauli paramagnetic parameter $\mu$ controls its strength.
In the rippled cylindrical Fermi surface, 
$\tilde{{\bf v}}_F\propto(\tilde{v}\cos\theta,\tilde{v}\sin\theta,\tilde{v}_z\sin k_z)$
with $\tilde{v}_z=0.5$ and $\tilde{v}=1+\beta\cos4\theta$
where $\beta$ signifies the Fermi velocity anisotropy.
The anisotropy ratio of the coherent lengths $\xi_c/\xi_{ab}\sim0.5$ by this $\tilde{v}_z$.
The notation and the procedure of the calculation are the same as before~\cite{ichioka}
albeit the field direction here is in the $ab$-plane.
After solving those equations self-consistently,
we evaluate the Helmholtz free energy and the entropy, which lead to the specific heat 
and the fourfold oscillation amplitude $A_4(T,H)$.

We have done extensive numerical computations of $A_4(T,H)$
for various $\mu$ and $\beta$ values, including the previous case with $\mu=0$
and $\beta=0$ whose result is basically consistent with that in Ref.\cite{vorontsov}.
Here we take into account the Pauli paramagnetic effect and the Fermi velocity anisotropy effect,
both of which are known to be important in CeCoIn$_5$.
In Fig. 4 we show the contour plot of $A_4(T,H)$ for $\mu=2$ and $\beta=0.5$.
It is seen that 
(i) The ADOS-quasiclassical region is confined to a narrow region in
low $H$ and $T$ bounded by $H_s$ and $T_s$. 
This region is characterized by $A_4<0$ where the oscillation sense coincides with 
the angle-resolved DOS, namely the maximum occurs when the field direction 
is along the antinodal [100] direction expected by $d_{x^2-y^2}$ symmetry.
(ii) The other region with $A_4>0$ in Fig. 4 is occupied by the reversed oscillation region.
The maximum oscillation is located around $T/T_{\rm c}\sim 0.25$ and $H/H_{c2}\sim 0.35$.
(iii) The overall landscape $A_4(T,H)$ well explains the experimental data in Fig. 3.
Thus we firmly conclude that the newly discovered sign reversal is understood
as $d_{x^2-y^2}$ symmetry.
(iv) We note the followings:
The ADOS-quasiclassical region in Fig. 4 ends at  $H_s/H_{c2}\sim0.3$
and $T_s/T_c\sim0.1$ while the experimental data show 
$H_s/H_{c2}\sim0.15$ and $T_s/T_c\sim0.1$.
There is a large discrepancy in $H_s$ value.
Note that $H_s/H_{c2}\sim0.4$ for the $d$-wave case with $\mu=\beta=0$
and $H_s/H_{c2}\sim0.35$ for the $d$-wave case with $\mu=2$ and $\beta=0$.
Thus the upper bound $H_s$ is sensitive for material parameters while $T_s/T_c$
is rather independent of those. This robustness is important when interpreting data,
namely to find the sign change the $T$ sweep is more effective than $H$ sweep.

\begin{figure}[t] 
\includegraphics[width=7.5cm, height=6 cm]{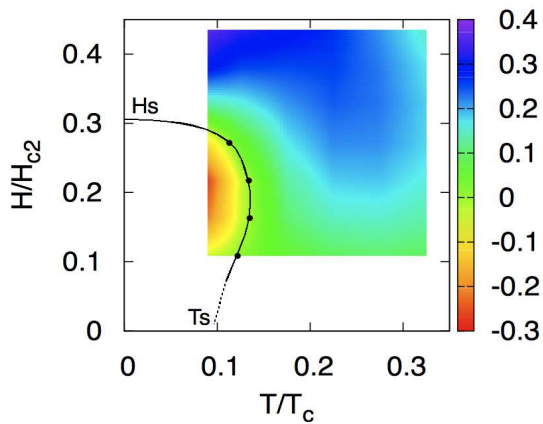}
  \caption{Topography of the specific heat oscillation amplitude
  $A_4$ (in units of percent) calculated under the paramagnetic effect parameter
  $\mu=2.0$,  the Fermi velocity anisotropy 
  $\beta=0.5$, and GL parameter  $\kappa=89$.
  The inside of the curved area around the origin
  is the ADOS-quasiclassical region bounded by $H_s$ and $T_s$.
  The calculations are done for $6\times7$
  grid points in $(T,H)$ and interpolated.
}
 \label{fig:fig4}
\end{figure}

Thermal conductivity experiment $\kappa_{\phi}$ by Izawa {\it et al.}~\cite{izawa} on CeCoIn$_5$
 is done outside the  ADOS-quasiclassical region ($T/T_{\rm c}\ge 0.2$).
 They correctly conclude the $d_{x^2-y^2}$ symmetry, which is fully 
 consistent with the present identification.
 We comment on the recent $C_{\phi}$ experiment by Park {\it et al.}~\cite{t-park08}
on CeRhIn$_5$ under pressure taken at $H/H_{c2}\sim 0.13$ and $T/T_{\rm c}=0.14$.
The data show the maximum at [110] direction.
Since that temperature is just at the boundary of the ADOS-quasiclassical region,
we should be careful to conclude the gap symmetry.
Based on the present result, we suggest an experiment by changing $T$
under fixed field to check whether the data is inside or outside that region.

In summary, we have found the sign change of the specific heat oscillation
and discovered the $(H,T)$ region, or ADOS-quasiclassical region 
where the oscillation maximum coincides
with the anti-nodal direction [100]. Except for trivial sign changes
near $H_{c2}$ (see for example \cite{deguchi}), this is the first time.
Thus the gap symmetry in CeCoIn$_5$ is $d_{x^2-y^2}$.
Those findings also help establishing the angle-resolved thermodynamic measurements
as a unique and indispensable spectroscopic method to 
identify the gap location, or the pairing symmetry.

\begin{acknowledgments}
This work has partly been supported by a Grant-in-Aid for Scientific
Research on Innovative Areas ``Heavy Electrons'' (No. 20102007) of the Ministry of Education, Culture, Sports and Technology, Japan.
\end{acknowledgments}
%
\bibliography{basename of .bib file}

\begin{thebibliography}{99}

\bibitem{tsuei}
C.C. Tsuei {\it et al}., Phys. Rev. Lett. {\bf 73}, 593 (1994); Nature {\bf 387}, 481 (1997).

 \bibitem{vekhter}
I. Vekhter, P.J. Hirschfeld, J.P. Carbotte, and E.J. Nicol, Phys. Rev.
 B{\bf 59}, R9023 (1999).
 
\bibitem{miranovic}
P. Miranovi\'{c}, M. Ichioka, K. Machida, and N. Nakai, J. Phys.: Condens. Matter
 {\bf 17}, 7971 (2005).
 
\bibitem{volovik}
G.E. Volovik, Pis'ma Zh. Eksp. Teor. Fiz. {\bf 58}, 457 (1993)
[JETP Lett. {\bf 58}, 469 (1993)].

\bibitem{t-park03}
T. Park, M.B. Salamon, E.M. Choi, H.J. Kim, and S.-I. Lee, Phys. Rev. Lett. {\bf 90}, 177001 (2003).

\bibitem{aoki}
H. Aoki {\it et al}., J. Phys.: Condens. Matter. {\bf 16}, L13 (2004).

\bibitem{deguchi}
K. Deguchi, Z.Q. Mao, H. Yaguchi, and Y. Maeno, Phys. Rev. Lett. {\bf 92}, 047002 (2004).

\bibitem{yamada}
A. Yamada {\it et al}., J. Phys. Soc. Jpn. {\bf 76}, 123704 (2007).

\bibitem{yano}
K. Yano {\it et al}., Phys. Rev. Lett. {\bf 100}, 017004 (2008).

\bibitem{t-park08}
T. Park, E.D. Bauer, and J.D. Thompson, Phys. Rev. Lett. {\bf 101}, 177002 (2008).
 
\bibitem{r-ikeda}
R. Ikeda and H. Adachi,  Phys. Rev. B {\bf 69}, 212506 (2004). 

\bibitem{izawa} 
K. Izawa {\it et al}., Phys. Rev. Lett. {\bf 87}, 057002 (2001).

\bibitem{bianchi}
A.D. Bianchi {\it et al}., Science {\bf 319}, 177 (2008). 

\bibitem{kawamura}
S.Ohira-Kawamura {\it et al.}, J. Phys. Soc. Jpn. {\bf 77}, 023702 (2008). 

\bibitem{hiasa}
N. Hiasa and R. Ikeda, Phys. Rev. Lett. {\bf 101}, 027001 (2008). 

\bibitem{machida}
K.M. Suzuki, M. Ichioka, and K. Machida, unpublished.

\bibitem{w-park}
W.K. Park, J.L. Sarrao, J.D. Thompson, and L.H. Greene, Phys. Rev. Lett. {\bf 100}, 177001 (2008). 

\bibitem{stock}
C. Stock, C. Broholm, J. Huidis, H.J. Kang, and C. Petrovic, Phys. Rev. Lett. {\bf 100}, 087001 (2008). 

\bibitem{eremin}
I. Eremin, G. Zwicknagl, P. Thalmeier, and P. Fulde, Phys. Rev. Lett. {\bf 101}, 187001 (2008).

\bibitem{vorontsov}
A.B. Vorontsov and I. Vekhter, Phys. Rev. B {\bf 75}, 224501 (2007). 

\bibitem{boyd}
G.R. Boyd,  P.J. Hirschfeld, I. Vekhter, and A.B. Vorontsov, Phys. Rev. B {\bf 79}, 064525 (2009).

\bibitem{movshovich}
R. Movshovich {\it et al}., Phys. Rev. Lett. {\bf 86}, 5152 (2001). 

\bibitem{s-ikeda}
S. Ikeda {\it et al}., J. Phys. Soc. Jpn. {\bf 70}, 2248 (2001).

\bibitem{nakai} 
For nodal SCs, $C/T$ at low $T$ is expected to be linear to $T$ even in magnetic fields: 
N. Nakai, P. Miranovi\'{c}, M. Ichioka, and K. Machida,
Phys. Rev. B {\bf 73}, 172501 (2006). 
 
\bibitem{curro}
N.J. Curro {\it et al}.,
Phys. Rev. B {\bf 64}, 180514(R) (2001).

\bibitem{ichioka}  
M. Ichioka and K. Machida, Phys. Rev. B {\bf 76}, 064502 (2007);
M. Ichioka, H. Adachi, T. Mizushima, and K. Machida, Phys. Rev. B {\bf 76}, 014503 (2007).



\end{thebibliography}

\end{document}